\documentclass[intlimits,twoside,a4paper]{article}

\usepackage[cp1251]{inputenc}

\usepackage{cmpj3}

\usepackage{bm}

\issue{2019}{22}{3}{33703}
\doinumber{10.5488/CMP.22.33703}

\title[Method of successive separation and summing of multiplicative diagrams of mass operator]%
{Method of successive separation and summing of multiplicative diagrams of mass operator for the multi-level quasiparticle interacting with polarization phonons}
\author[M.V.~Tkach, Ju.O.~Seti, O.M.~Voitsekhivska, V.V.~Gutiv]{M.V.~Tkach\footnote{E-mail: m.tkach@chnu.edu.ua},
Ju.O.~Seti, O.M.~Voitsekhivska, V.V.~Gutiv}
\address{Yuriy Fedkovych Chernivtsi National University, 2 Kotsyubinsky St., 58012 Chernivtsi, Ukraine}

\date{Received	June 17, 2019, in final form August 20, 2019}

\begin{document}

\maketitle

\begin{abstract}
The theory of renormalized energy spectrum of a multi-level quasiparticle interacting with polarization phonons at $T=0$~K is developed within the Feynman-Pines diagram technique in a new approach. It permits a successive separation of multiplicative diagrams from non-multiplicative ones for all orders of mass operator, and their partial summing as well. The obtained mass operator is presented as a sum of branched chain fractions, which effectively take into account the multi-phonon processes. For the two-level quasiparticle it is shown that just the interlevel (non-diagonal) interaction with phonons fundamentally changes the properties of the spectrum. In the vicinity of all threshold energies, the quasi-equidistant phonon satellite bands (i.e., groups of energy levels) are formed. They correspond to the complexes of bound states of a quasi-particle with many phonons.
\pacs 71.38.-k, 72.10.Di, 63.20.kk
\end{abstract}

\section{Introduction}
Recently, new physical phenomena (in particular, high temperature superconductivity) in 3D and low-dimensional structures associated with the effects of spatial quantization have greatly urged an intensive development of new theoretical approaches to the study of interaction of quasiparticles (electron, hole, exciton) with different types of phonons. All these various of  interactions play an important role in the formation and in temperature evolution of the renormalized spectra and, hence, in the observed physical phenomena.

Generally, during the recent two decades, new powerful and phighly promissing methods have been scrupulously developed in order to calculate the spectra of quasiparticles renormalized due to interaction with phonons in polaron models (Fr\"ohlich, Holstein, Su-Schrieffer-Heeger) which have different spatial dimension. The progress in the solution of polaron problems has been reached due to the use of modern analytical and analytically-numerical methods: exact diagonalization  (ED) \cite{Ale94}, modern variational method (VM) \cite{Ku02,Hir82}, dynamical mean-field theory (DMFT) \cite{Ciu97}, density-matrix renormalization group (DMRG) \cite{Zha99,Kor99}, quantum Monte Carlo (QMC) \cite{Spe05,Hoh05,Hag06}, diagrammatic Monte Carlo (DMC) \cite{Pro98,Mis00,Mis01,Gou16,Mis18}, the bold diagrammatic Monte Carlo (BDQMC) \cite{Pro07,Hou12,Kul13,Mis14}, the momentum average approximation (MA) \cite{Ber06,Mar10,Ebr12,Mol16,Mar17}.

The results presented in the papers where these non-perturbed methods have been used, proved to be well correlating for different models of the structures where the interaction between quasiparticles and phonons is important in a wide range of their coupling (from weak to strong). In the majority of papers, the one-band models for quasiparticles were usually studied. However, recently, the systems of quasiparticles with a more complicated spectrum were observed \cite{Mol16,Mar17}. The arising interest to the multi-level (multi-band) quasiparticles interacting with phonons is caused by the rapid development of physics of nanoheterostructures, where the energy spectrum of even uncoupled quasiparticles is multi-level (multi-band) due to the effect of spatial quantization.

The theory of renormalized energy spectrum of multi-band quasiparticles interacting with different phonon modes (confined, interface, propagating and others) \cite{Str01,Zhu12,Zha13,Set18} in nanoheterostructures is not only of theoretical importance but it has an essential practical significance. Although the multilayered nanostructures have been long used  as components of unique modern devices [quantum cascade lasers (QCL) and detectors (QCD)] operating in the actual infra-red range of electromagnetic waves, the consistent theory of their physical phenomena is far from being complete.

It is well known \cite{Hof08,Lei08,Gio09,Hof10} that phonons play an important positive role for the QCDs functioning in the middle and in the far IR-range because their extractors operate within the so-called ``complete phonon ladder''. The quantum wells and barriers of the extractors have got the composition and design of sizes which provide the needed number of equidistant electron energy levels (with the distance of one phonon energy) that create this ``ladder''. Herein, the tunneling electron performs consistent radiationless one-phonon quantum transitions from the upper level of the cascade active region exactly to the lower (ground) level of the active region of the next cascade through this ladder, which ensures an appropriate QCD operation.

The described mechanism of extractor operating within the ``complete phonon ladder'', which relaxes the electron energy between the active regions of the neighbouring cascades, is not the only one. Recently, it turned out that the cascades of newly produced and appropriately operating QCD of near IR-range \cite{Var08,Sak12,Sak13,Bee13} work with the extractors having a ``torn phonon ladder'', lacking almost a half of the lower energy levels! A detailed review and analysis of experimental papers show that in some QCD operating in the other ranges \cite{Rei14,Rei15} the extractors were imperfect with respect to the equivalent distance between the levels that create a ``phonon ladder''.

There is still no theory of electron-phonon interaction that would reliably describe the physical phenomena in multilayered nanostructures. However, the satellite states with equidistant energy spectrum arising due to the interaction between one-level quasiparticles and phonons were studied in theoretical papers \cite{Dav76,Lev74,Tka84}. It gives reason to expect that multi-level uncoupled electrons interacting with phonons in quasi-two-dimensional multilayered nanostructures are capable of creating quasi-equidistant satellite mini-bands, which would compensate both for the energy levels lacking in the ``phonon ladder'' and for their imperfect equidistant characteristics.

For the simplest model of an extractor with the ``torn phonon ladder'', created by two energy levels of a quasiparticle and by their phonon satellites, in the papers \cite{Tka18,Tka19} there was presented a theory of an energy spectrum of two-level localized quasiparticle interacting with polarization phonons at $T=0$~K. Using the modified method of Feynman-Pines diagram technique, it was shown that in the first approximation of renormalized mass operator (MO) (where all multiplicative diagrams are taken into account completely), the satellite mini-bands are really observed in the spectrum.

In this paper, we develop a generalized theory for the spectrum of a multi-level quasiparticle re\-nor\-ma\-lized due to the interaction with phonons, taking into account not only the main multiplicative diagrams in the mass operator but also the diagrams that are successively separated from its non-multiplicative block. For the same model as in \cite{Tka18,Tka19}, it is shown that the renormalized spectrum of the system, apart from the known properties, also prossesses some new important ones.

\section{Hamiltonian of the system. Mass operator of quasiparticle Green's function at $T=0$~K}

To obtain a renormalized spectrum of the system consisting of a multi-level localized quasiparticle interacting with polarization phonons, we use the Fr\"ohlich Hamiltonian, like in \cite{Tka18,Tka19,Lev74}
\begin{equation} \label{Eq1}
\hat{H}=\sum _{\mu =1}^{\tau } E_{\mu } a_{\mu }^{+}  a_{\mu } +\sum _{\vec{q}}\Omega (\vec{q})\left(b_{\vec{q}}^{+} b_{\vec{q}} +\frac{1}{2} \right) +\sum _{\vec{q}} \sum_{\mu_{1} ,\mu _{2} =1}^{\tau} \varphi_{\mu_{1} \mu_{2}}  a_{\mu_{1} }^{+} a_{\mu_{2} } \big(b_{\vec{q}} +b_{-\vec{q}}^{+} \big).
\end{equation}
Here, $E_{\mu=1,\ldots,\tau}$ are the energy levels of an uncoupled quasiparticle. As the energy of optical phonon typically weakly depends on its quasimomentum ($\vec{q}$), we further put $\Omega(\vec{q})=\Omega$, neglecting the dispersion. Coupling constants ($\varphi_{\mu_{1}\mu_{2}}$), which describe quasiparticle-phonons interaction, are assumed to be the known parameters and characterize either intra-level interaction at $\mu_{1}=\mu_{2}$ or the inter-level interaction at $\mu_{1}\neq \mu_{2}$.

We should note that Hamiltonian like~(\ref{Eq1}) can describe a wide range of 3D-models (for example, impurity centers) and spatially confined 1D- or 2D-models of low dimensional structures containing multi-level localized quasiparticles interacting with confined or interface phonons.

The energy spectrum of the system renormalized due to the interaction at cryogenic temperature (formally $T=0$~K) is obtained within the method of Feynman-Pines diagram technique \cite{Tka19,Fey62,Abr12} for the Fourier image of quasiparticle casual Green's function $G_{\mu \mu '}(\omega)$. Moreover, we use the approach proposed in papers \cite{Tka18,Tka19}, modified for the case of a quasiparticle [with an arbitrary number ($\tau$) of levels] interacting with phonons. Taking into account the Hamiltonian~(\ref{Eq1}), the Green's functions $G_{\mu \mu '}(\omega)$ satisfy the system of $\tau^{2}$ Dyson equations
\begin{equation} \label{Eq2}
G_{\mu \mu '} (\omega )=(\omega -E_{\mu } +i\eta )^{-1} \bigg[\delta _{\mu \mu '} +\sum _{\mu _{1}=1}^{\tau } M_{\mu \mu_{1} }(\omega ) G_{\mu_{1}\mu'} (\omega)\bigg],\, \, \, \, \, \mu,\, \mu '=1,\, 2,\, \ldots ,\tau ;\, \, \, \, (\eta \to 0),
\end{equation}
where $M_{\mu \mu_{1}}$ is the complete matrix MO and $E_{\mu=2,\ldots,\tau}=E_{1}+\Delta E_{\mu=2,\ldots,\tau}$. This system is conveniently solved in dimensionless functions, variables and constants:
\begin{equation} \label{Eq3}
\begin{array}{c} {g_{\mu}(\xi)=g_{\mu \mu}(\xi )=\Omega G_{\mu \mu}\,,\qquad m_{\mu}(\xi )=M_{\mu} \Omega ^{-1} ,\qquad m_{\mu \mu} (\xi)=M_{\mu \mu} \Omega ^{-1},} \\
\\
{\xi =(\omega-E_{1})\Omega^{-1},\, \, \, \, \, \xi _{\mu _{l}} =(\omega -E_{\mu _{l} } )\Omega ^{-1}, \, \, \, \, \, \alpha _{\mu_{1} \mu_{2}} =\varphi_{\mu_{1} \mu_{2}} \Omega ^{-1}, \, \, \, \, \, \delta_{\mu =2,\, \ldots ,\tau} =\Delta E_{\mu =2,\, \ldots ,\tau} \Omega ^{-1}. }
\end{array}
\end{equation}

The system of $\tau^{2}$ dimensionless equations obtained from~(\ref{Eq2}) is sophisticated (for big $\tau$ numbers), though it is solved exactly. For the dimensionless Green's function of $\mu$-th level [$g_{\mu}(\xi)=g_{\mu\mu}(\xi)$], the solution is defined by Dyson equation (at $\hbar=1$)
\begin{equation} \label{Eq4}
g_{\mu}(\xi)=\{ \xi_{\mu} - m_{\mu}(\xi)\}^{-1}
\end{equation}
within  the complete MO $m_{\mu}(\xi)$, which can be written in the form
\begin{equation} \label{Eq5}
m_{\mu}(\xi)=m_{\mu \mu}(\xi)+m_{\mu \mu}^{g}(\xi) \, ,
\end{equation}
where $m_{\mu\mu}$ is a diagonal component, which describes all intra-level interactions with phonons but without taking into account non-diagonal elements of $g_{\mu\mu'}$ matrix. The components $m_{\mu\mu}^{g}$, produced by non-diagonal elements of $g_{\mu\mu'}$, are defined exactly. However, they have a sophisticated structure for big $\tau$ numbers, presented, for example, in appendix~\ref{appendix_A} at $\tau=3,4,5$. For the two-level system ($\tau=2$), the diagonal elements are given by simple formulae
\begin{equation} \label{Eq6}
m_{11}^{g(2)}(\xi)=\frac{m_{12}(\xi) m_{21}(\xi)}{\xi_{2} - m_{22}(\xi)} \, ,\qquad \qquad m_{22}^{g(2)}(\xi)=\frac{m_{21}(\xi) m_{12}(\xi)}{\xi_{1} - m_{11}(\xi)} \, .
\end{equation}

The matrix of complete MO $m_{\mu\mu'}$ is defined by the rules of Feynman-Pines diagram technique \cite{Abr12}, which are generalized for the case of multi-level systems. Since the energies of quasiparticle and phonons are dispersionless, the equivalent diagrams with and without crossing the phonon lines, correspond to the identical analytical expressions, such as
\begin{equation} \label{Eq7}
\begin{array}{c}
\includegraphics[width=2.2in, keepaspectratio=true]{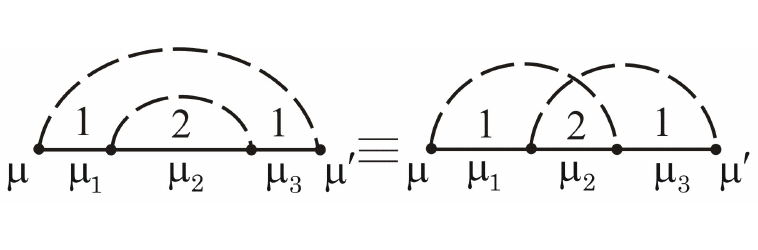}\raisebox{12pt}{$\hspace{1mm}.$} 
\end{array}
\end{equation}
Thus, contrary to the classic Fr\"ohlich one-band Hamiltonian \cite{Tka19,Abr12}, where the energy of an uncoupled quasiparticle is the function of quasimomentum, the dimensionless MO $m_{\mu\mu'}$ is obtained in such diagrammatic representation that contains only all non-equivalent diagrams without crossing the phonon lines. The number of equivalent diagrams of this type is fixed by the  integers prior to the respective diagrams.
\begin{equation} \label{Eq8}
\includegraphics[width=5.7in, keepaspectratio=true]{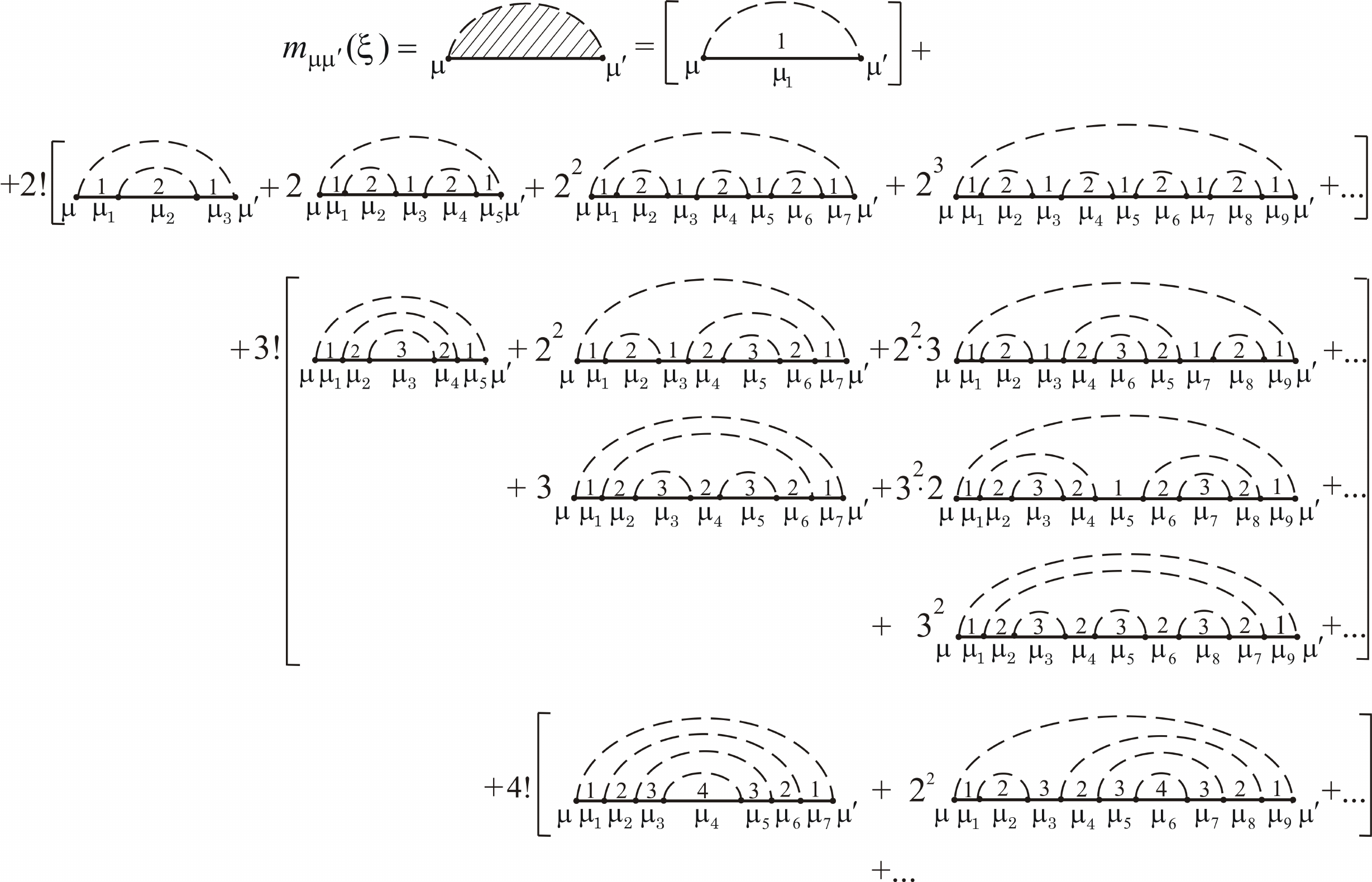}
\end{equation}

All MO diagrams in an arbitrary $p$-th order over the power of the pairs of coupling constants, which are the same as the number of phonon lines in a respective diagram, are definitely calculated by a computer program. It would be clear further on that in order to establish the exact rules of partial summing of infinite ranges of diagrams, it is quite sufficient to take into account all the first 128 non-equivalent diagrams to the eighth order inclusive.

The analytical expression for an arbitrary diagram is obtained as the sum over all inner indices ($\mu_{1}, \mu_{2}, \ldots, \mu_{N}$), except the outer ones ($\mu, \mu'$), of the products of all vertices and solid lines
\begin{equation} \label{Eq9}
\begin{array}{c}
\hfil \includegraphics[width=1.5in, keepaspectratio=true]{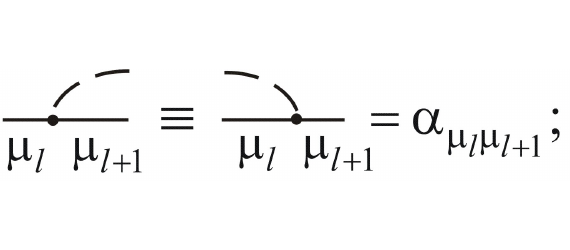} \qquad \qquad \qquad  \includegraphics[width=1.5in, keepaspectratio=true]{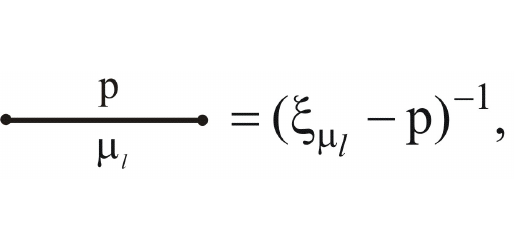}
\end{array}
\end{equation}
where $p$ is a number of dashed (phonon) lines situated above the solid (quasiparticle) line with index $\mu_{l}$. For example, the first two components of $m_{\mu\mu'}$ are of the following diagrammatic and analytical forms:
\begin{equation} \label{Eq10}
\begin{array}{c}
\includegraphics[width=2.3in, keepaspectratio=true]{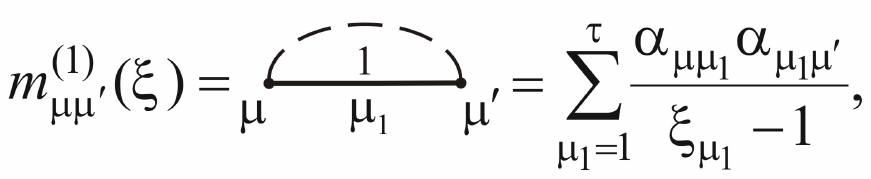}
\end{array}
\end{equation}
\begin{equation} \label{Eq11}
\begin{array}{c}
\includegraphics[width=4in, keepaspectratio=true]{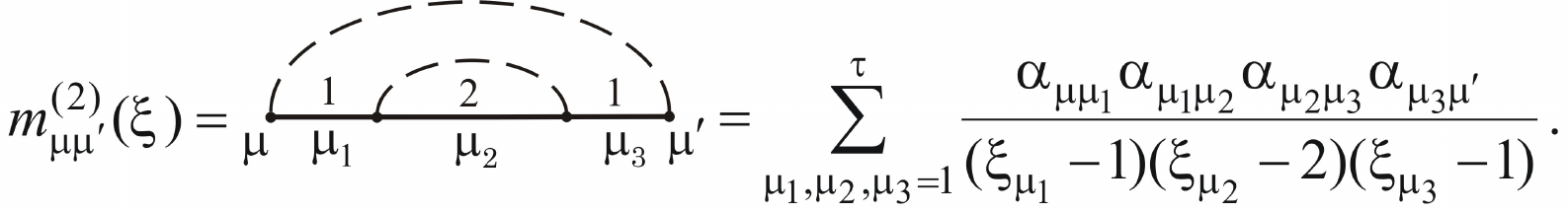}
\end{array}
\end{equation}

The renormalized spectrum of the system is calculated in a wide energy range, containing not only the ground level but the levels of the excited states, which arise in the vicinity of the thresholds of the radiation of phonons. Thus, it is necessary to represent the complete MO~(\ref{Eq8}) in the form which effectively takes into account both the inter-level interaction and multi-phonon processes. However, the classic form~(\ref{Eq8}) does not satisfy the second condition, in particular, due to the so-called ''problem of sign''. In order to solve it, one should perform a successive partial summing of infinite ranges of diagrams in the complete MO.

\section{Method of successive partial summing of MO diagrams}

To perform the partial summing of diagrams in the complete MO $m_{\mu\mu'}$, it is convenient to group them into the classes which are written in brackets in expression~(\ref{Eq8}). It is clear that the $p$-th class of renormalized diagrams (except the first one) together with the factor $p!$ contains an infinite number of only those diagrams (together with numerical factors), whose arbitrary blocks, in their turn, contain not more than $p$ dashed lines over any of the solid ones. This class of diagrams is further referred to as a renormalized $p$-phonon MO and is denoted by $m_{\mu\mu'}^{[p]}$. We should note that, as it is clear from~(\ref{Eq8}), only one-phonon diagram~(\ref{Eq10}) in its first class of partial $p$-phonon diagrams is single, hence, $m_{\mu\mu'}^{[1]}=m_{\mu\mu'}^{(1)}$. All other classes of renormalized MO $\big(m_{\mu\mu'}^{[p \geq 2]}\big)$ are given by the respective infinite ranges of diagrams. For example, the renormalized two-phonon MO is
\begin{equation} \label{Eq12}
\begin{array}{c}
\includegraphics[width=3.5in, keepaspectratio=true]{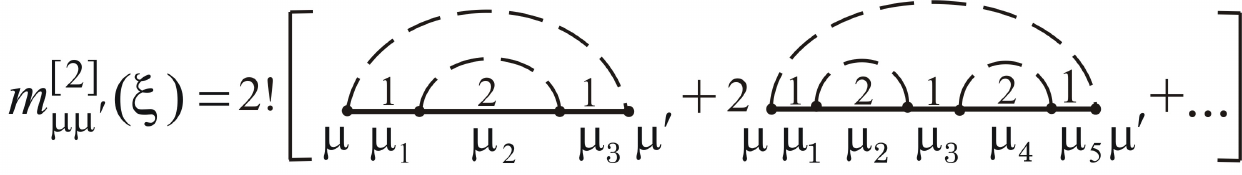}\raisebox{12pt}{$\hspace{1mm}.$}
\end{array}
\end{equation}
Then, the complete MOs $m_{\mu\mu'}$ can be written as a sum of renormalized MO of all orders
\begin{equation} \label{Eq13}
m_{\mu \mu '} (\xi)=\sum _{p=1}^{\infty} m_{\mu \mu '}^{[p]}(\xi ) \, .
\end{equation}

In each MO diagram, the analytical contributions of its elements are summed up over all $\mu_l$ indices, except the first one ($m_{\mu\mu'}^{[1]}$). Hence, none of them is multiplicative, in general form. However, the analytical expressions show that each diagram can be expressed as a sum of two components: multiplicative (m) and non-multiplicative (nm). To this end, in any diagram, in which summing is performed over all inner indices [for example, $\mu_{1},~\mu_{2},~\mu_{3}$ in the first diagram of~(\ref{Eq12})], one should separate the multiplicative component having certain equal indices (for example, $\mu_{1},~\mu_{2},~\mu_{3}=\mu_{1}$), where the summing is performed only over the indices $\mu_{1},~\mu_{2}$ and the non-multiplicative component having the indices ($\mu_{1},~\mu_{2},~\mu_{3}\neq\mu_{1}$), where the summing is performed over all these indices. In this approach, the non-renormalized MO of the second order over the number of phonon lines is expressed in the following diagrammatic and analytical forms:
\begin{equation} \label{Eq14}
\begin{array}{c}
\includegraphics[height=1.0in, keepaspectratio=true]{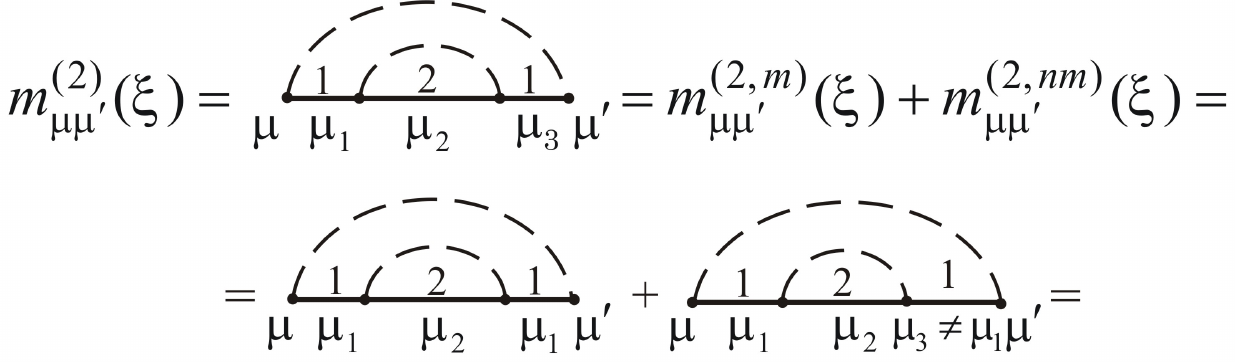}\\
\displaystyle =\sum _{\mu _{1} =1}^{\tau }\frac{\alpha _{\mu \mu _{1}} m_{\mu _{1} \mu _{1}}^{(1)} (\xi -1)\, \alpha _{\mu _{1} \mu '}}{(\xi_{\mu _{1}} -1)^{2}}  +\sum _{\mu _{1} =1}^{\tau}\, \frac{\alpha _{\mu \mu _{1}}}{\xi _{\mu _{1}} -1}  \, \sum _{\mu _{3} \ne \mu _{1} }^{\tau }\frac{m_{\mu _{1} \mu _{3} }^{(1)} (\xi -1)\, \alpha _{\mu _{3} \mu '} }{\xi _{\mu _{3} } -1} \, .
\end{array}
\end{equation}
The non-renormalized MO of the third order is as follows:
\begin{equation} 
\includegraphics[height=1.0in, keepaspectratio=true]{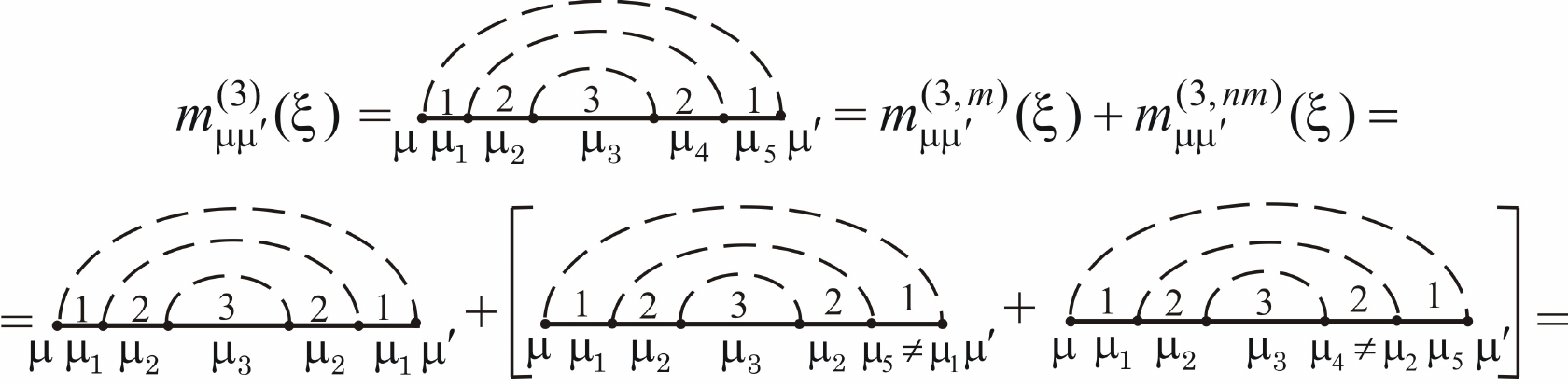} \nonumber
\end{equation}
\begin{equation} \label{Eq15}
\begin{array}{c}
\displaystyle  =\sum _{\mu _{1}, \mu _{2} =1}^{\tau } \frac{\alpha _{\mu \mu _{1}} \alpha _{\mu_{1} \mu _{2}} m_{\mu _{2} \mu _{2}}^{(1)} (\xi -2) \, \alpha_{\mu _{2} \mu _{1}} \alpha _{\mu _{1} \mu '}}{(\xi _{\mu _{1}} -1)^{2} (\xi _{\mu _{2}} -2)^{2} }  + \left[\sum _{\scriptsize{\begin{array}{l} {\mu _{1}, \mu_{2},}\\{ \mu _{5} \ne \mu _{1}}\end{array}}}^{\tau } \frac{\alpha _{\mu \mu _{1}} \alpha _{\mu _{1} \mu _{2}} m_{\mu _{2} \mu _{2}}^{(1)} (\xi -2) \, \alpha_{\mu _{2} \mu _{5}} \alpha _{\mu _{5} \mu '}}{(\xi _{\mu _{1}} -1)(\xi _{\mu _{2}} -2)^{2} (\xi _{\mu _{5}} -1)} + \right.\\
\displaystyle \left. + \sum_{\scriptsize{\begin{array}{l} {\mu _{1}, \mu _{2}, \mu _{5},}\\{ \mu _{4} \ne \mu _{2}} \end{array}}}^{\tau } \frac{\alpha _{\mu \mu _{1}} \alpha _{\mu _{1} \mu _{2}} m_{\mu _{2} \mu _{4}}^{(1)} (\xi -2)\, \alpha _{\mu _{4} \mu _{5}} \alpha _{\mu _{5} \mu '}}{(\xi _{\mu _{1}} -1)(\xi _{\mu _{2}} -2)(\xi _{\mu _{4}} -2)(\xi _{\mu _{5} } -1)}  \right].
\end{array}
\end{equation}

Having separated all the diagrams of all orders at m- and nm- classes in the complete MO~(\ref{Eq8}) and taking into account that one-phonon MO $m_{\mu\mu'}^{(1)}$ belongs to the m-class of diagrams, the complete MO~(\ref{Eq13}) can be written as a sum of two components
\begin{equation} \label{Eq16}
m_{\mu \mu '} (\xi) = m_{\mu \mu '}^{1,m}(\xi) + m_{\mu \mu '}^{2,nm}(\xi) = \sum _{p=1}^{\infty} m_{\mu \mu '}^{p,[m]}(\xi) + \sum _{p=2}^{\infty} m_{\mu \mu '}^{p,[nm]}(\xi) \, .
\end{equation}
The first component ($m_{\mu\mu'}^{1,m}$) contains one-phonon diagram renormalized by all m-diagrams of all orders and the second one ($m_{\mu\mu'}^{2,nm}$) contains the rest nm-diagrams of all orders beginning from the second, in which one should further successively separate other classes of m- and nm-diagrams of higher orders over the powers of coupling constants (phonon lines). The multiplicative structure of all terms in MO $m_{\mu\mu'}^{1,m}$ makes it possible to perform an exact partial summing. This was described in detail in papers \cite{Tka18,Tka19}, where the latter was denoted as $m_{\mu\mu'}^{[m]}$.

As a result, being renormalized by all the multiphonon process, one-phonon MO in the first approximation $m_{\mu\mu'}^{1,m}$ is obtained in the form of infinite branched chain fraction
\begin{equation} \label{Eq17}
\begin{array}{c}
\includegraphics[height=1.0in, keepaspectratio=true]{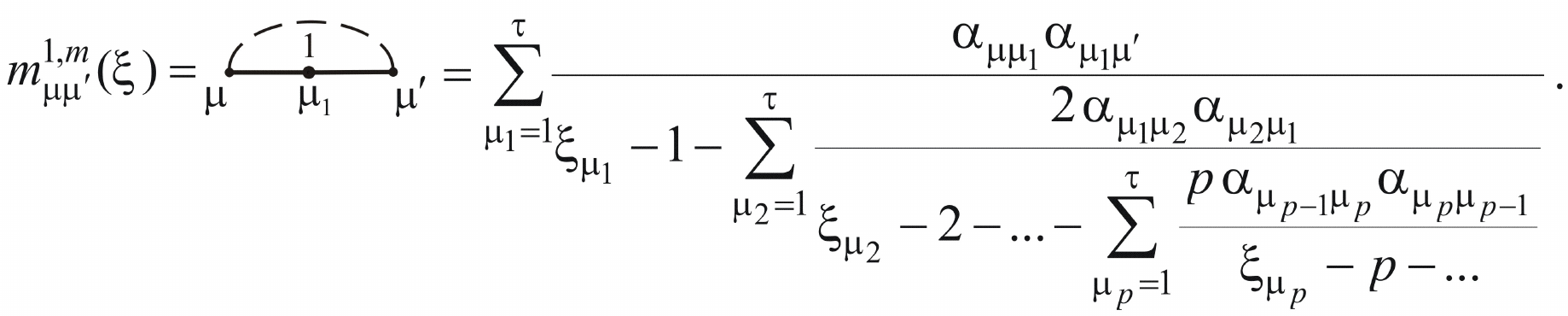}
\end{array}
\end{equation}
For further calculations it is convenient  to represent the latter expression as identical functional equation
\begin{equation} \label{Eq18}
m_{\mu \mu '}^{1,m}(\xi) = \sum _{\mu _{1}=1}^{\tau}\frac{\alpha _{\mu \mu _{1}} \alpha _{\mu _{1} \mu '}}{\xi _{\mu _{1}} -1-m_{\mu _{1} \mu _{1}}^{2,m}(\xi)} \, .
\end{equation}
Thus, for an arbitrary $p=1,2,\ldots, \infty$, the relationship is valid:
\begin{equation} \label{Eq19}
\displaystyle m_{\mu_{p-1} \mu'_{p-1}}^{p,m}(\xi) = \sum _{\mu_{p}=1}^{\tau} \frac{p\alpha_{\mu_{p-1} \mu_{p}} \alpha_{\mu_{p} \mu'_{p-1} }}{\xi_{\mu _{p}} -p-m_{\mu _{p} \mu _{p}}^{p+1,m}(\xi)} \, .
\end{equation}

The next stage in calculating the complete MO ($m_{\mu\mu'}$) is to separate all m-diagrams from the whole infinite number of nm-diagrams $m_{\mu\mu'}^{2,nm}$, starting from the two-phonon \includegraphics*[width=0.7in, height=0.2in, keepaspectratio=false]{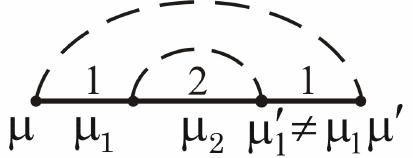}, which were not logged in $m_{\mu\mu'}^{1,m}$ and are renormalized due to all multi-phonon processes by two-phonon MO $m_{\mu\mu'}^{2,[m]}$ (being the correction of the second approximation for the complete MO). Then, all other diagrams of the complete MO, which were not logged in $m_{\mu\mu'}^{1,m}$  and  $m_{\mu\mu'}^{2,[m]}$, starting from the three-phonon ones, produce the non-multiplicative MO $m_{\mu\mu'}^{3,[nm]}$. As for $m_{\mu\mu'}^{2,[m]}$, it is partially summed up exactly in the same way as $m_{\mu\mu'}^{1,m}$. Thus, having separated from $m_{\mu\mu'}^{2,nm}$ the infinite ranges of only m-diagrams of the second approximation
\begin{equation} \label{Eq20}
\includegraphics[height=1.73in, keepaspectratio=true]{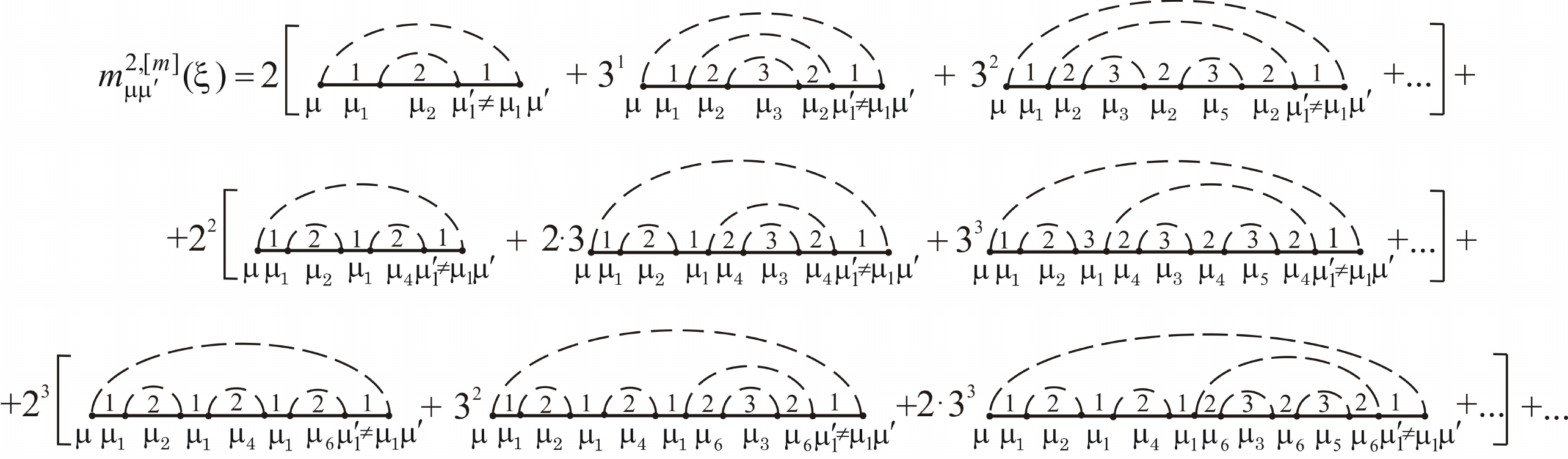}
\end{equation}
and having performed their exact partial summing according to the presented rules of diagram technique~(\ref{Eq9}), the following analytical expression is obtained
\begin{equation} \label{Eq21}
m_{\mu \mu'}^{2,[m]}(\xi) = \sum_{\scriptsize{\begin{array}{l} {\mu_{1}, \mu_{2},}\\{\mu'_{1} \ne \mu_{1}} \end{array}}}^{\tau} \frac{2 \alpha_{\mu \mu _{1}} \alpha_{\mu_{1} \mu_{2} } \alpha_{\mu_{2} \mu'_{1}} \alpha_{\mu'_{1} \mu'}}{(\xi_{\mu_{1}} -1)(\xi_{\mu'_{1}} -1)[\xi_{\mu_{2}} -2-m_{\mu_{2} \mu_{2}}^{3,m} (\xi)]} \sum_{s=0}^{\infty} \bigg\{\sum_{\mu'_{2}}^{\tau} \frac{2 \alpha_{\mu_{1} \mu'_{2}} \alpha_{\mu'_{2} \mu_{1}}}{(\xi_{\mu_{1}} -1)(\xi_{\mu_{2}} -2) \Big [1-\frac{m_{\mu_{2} \mu_{2}}^{3,m} (\xi)}{(\xi_{\mu_{2}} -2)} \Big]}  \bigg\}^{S},
\end{equation}
Partially summing the ranges of m-diagrams or, which is the same, converting the last row of geometric progression in~(\ref{Eq21}), final diagrammatic and analytical expressions for a completely renormalized (due to multi-phonon processes) two-phonon MO  $m_{\mu\mu'}^{2,[m]}$ (in the second approximation) are obtained
\begin{equation} \label{Eq22}
\begin{array}{c}
\includegraphics[height=0.45in, keepaspectratio=true]{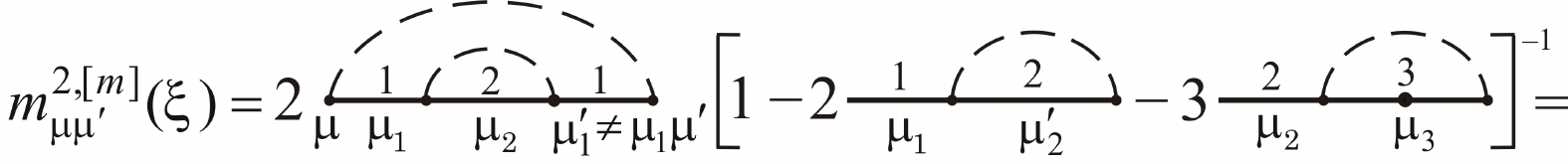}\\
\displaystyle =\sum _{\mu _{1}, \mu _{2} ,\mu '_{1} \ne \mu _{1}}^{\tau }\frac{2\alpha _{\mu \mu _{1} } \alpha _{\mu _{1} \mu _{2} } \alpha _{\mu _{2} \mu '_{1} } \alpha _{\mu '_{1} \mu '} }{(\xi _{\mu _{1} } -1)(\xi _{\mu _{2} } -2)(\xi _{\mu '_{1} } -1)\left[1-\frac{2\, m_{\mu _{1} \mu _{1} }^{(1)} (\xi -1)}{\xi _{\mu _{1} } -1} -\frac{m_{\mu _{2} \mu _{2} }^{3,m} (\xi )}{\xi _{\mu _{2} } -2} \right]} \, .
\end{array}
\end{equation}

In the same way, final diagrammatic and analytical expressions for the completely renormalized (due to multi-phonon processes) three-phonon MO $m_{\mu\mu'}^{3,[m]}$ (in the third approximation) are obtained
\begin{equation} \label{Eq23}
\begin{array}{c}
\includegraphics[height=1.05in, keepaspectratio=true]{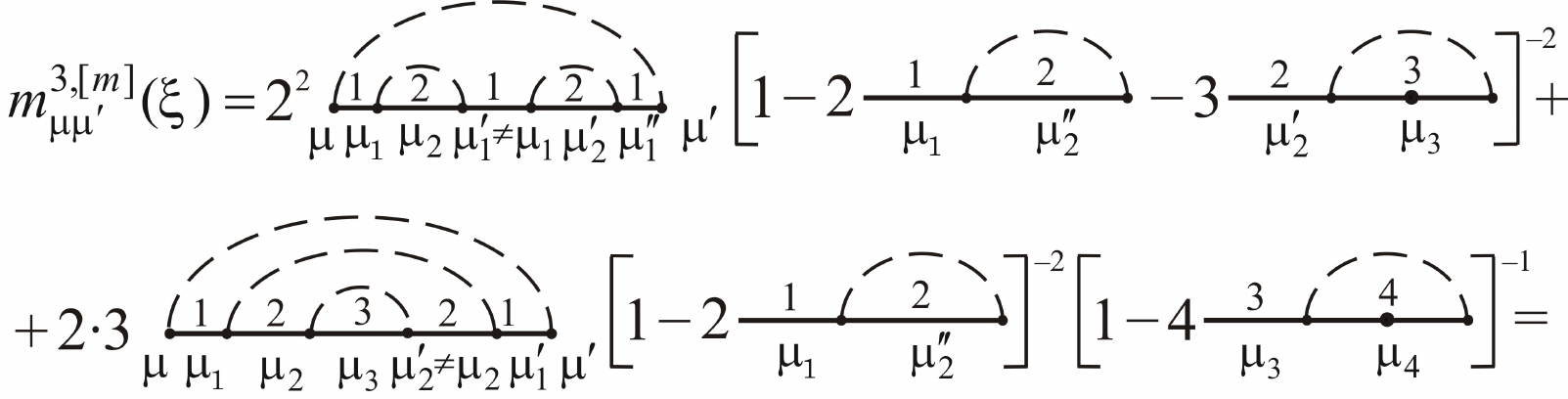}\\
\displaystyle {=2^{2} \sum_{\scriptsize{\begin{array}{l} {\mu_{1}, \mu_{2}, \mu'_{2},} \\ {\mu''_{1}, \mu'_{1} \ne \mu_{1}} \end{array}}}^{\tau} \frac{\alpha_{\mu \mu_{1}} \alpha_{\mu_{1} \mu_{2}} \alpha_{\mu_{2} \mu'_{1}} \alpha_{\mu'_{1} \mu'_{2}} \alpha_{\mu'_{2} \mu''_{1}} \alpha_{\mu''_{1} \mu'}}{(\xi_{\mu_{1}} -1)(\xi_{\mu_{2}} -2)(\xi_{\mu'_{1}} -1)(\xi_{\mu'_{2}} -2)(\xi_{\mu''_{1}} -1)}  \left[1-\frac{2\, m_{\mu_{1} \mu_{1}}^{(1)} (\xi -1)}{\xi_{\mu_{1}} -1} -\frac{m_{\mu'_{2} \mu'_{2}}^{3,m} (\xi)}{\xi_{\mu'_{2}} -2} \right]^{-2} +} \\
\displaystyle {+2 \cdot 3\sum_{\scriptsize{\begin{array}{l} {\mu_{1}, \mu_{2}, \mu_{3},} \\ {\mu'_{1}, \mu'_{2} \ne \mu_{2}} \end{array}}}^{\tau} \frac{\alpha_{\mu \mu_{1}} \alpha_{\mu_{1} \mu_{2}} \alpha_{\mu_{2} \mu_{3}} \alpha_{\mu_{3} \mu'_{2}} \alpha_{\mu_{2} \mu'_{1}} \alpha_{\mu'_{1} \mu'}}{(\xi_{\mu_{1}} -1)(\xi _{\mu _{2} } -2)(\xi _{\mu _{3} } -3)(\xi _{\mu '_{2} } -2)(\xi _{\mu '_{1} } -1)}  \left[1-\frac{2\, m_{\mu '_{1} \mu _{1} }^{(1)} (\xi -1)}{\xi_{\mu'_{1}} -1} \right]^{-2} \left[1-m_{\mu_{3} \mu_{3}}^{4,m} (\xi)\right]^{-1}}.
\end{array}
\end{equation}

A similar cyclic procedure of successive separation of m-diagrams from nm-diagrams with their subsequent summing up in the proposed way is totally applicable in all higher orders for MO. It is clear that with each next cycle, the number ($s$) of terms $m_{\mu\mu'}^{s,[m]}$ renormalized by m-diagrams and, consequently, their contribution into $m_{\mu\mu'}$ will be bigger while the number of terms $m_{\mu\mu'}^{s,nm}$ produced by nm-diagrams and their contribution will become smaller.

In the limiting case of infinite number of cycles ($s \rightarrow \infty$), $m_{\mu\mu'}^{s,nm} \rightarrow 0$. Thus, in $m_{\mu\mu'}$  there remains only the sum of all terms $m_{\mu\mu'}^{s,m}$ produced by all m-diagrams. Now it is possible to formally present the complete MO in the form
\begin{equation} \label{Eq24}
m_{\mu \mu'} (\xi)={\lim \limits_{s_{\max} \to \infty}} \sum_{s=1}^{s_{\max}}\left[m_{\mu \mu'}^{s,[m]}(\xi)+m_{\mu \mu'}^{s,nm}(\xi)\right] = \sum_{s=1}^{\infty}m_{\mu \mu'}^{s,[m]}(\xi) \, ,
\end{equation}
where, in order to unambiguously understand this formula, we use the denotion $m_{\mu\mu'}^{1,[m]}\equiv m_{\mu\mu'}^{1,m}$ at $s=1$.

The algorithm of successive separation of m-diagrams from the nm-diagrams with their partial summing is completely clear but if the number of terms ($s$) increases, the analytical expressions for $m_{\mu\mu'}^{s,[nm]}$ become more sophisticated. Therefore, when MO~(\ref{Eq24}) is practically used for the calculation of renormalized spectrum of the system, one should take into account this number of terms ($s$) in $m_{\mu\mu'}^{s,[nm]}$, which  provides sufficient convergence (saturation) of the spectrum.

In our previous paper \cite{Tka18}, we studied only the first ($m_{\mu\mu'}^{1,m}$) approximation for a complete MO and revealed the main property of a renormalized spectrum of two-level localized quasiparticle interacting with phonons, in a wide range of energies. In this paper, using the obtained MO in third approximation
\begin{equation} \label{Eq25}
m_{\mu \mu'}^\text{III}(\xi)=m_{\mu \mu'}^{1,[m]}(\xi)+m_{\mu \mu'}^{2,[m]}(\xi)+m_{\mu \mu'}^{3,[m]}(\xi)
\end{equation}
we shall clarify some new properties of this spectrum, which were not revealed in the first paper.

\section{Properties of renormalized energy spectrum of the system at $T=0$~K taking into account the higher successive approximations of mass \\ operator}
In our paper \cite{Tka18}, we studied the renormalized spectrum of a two-level quasiparticle interacting with dispersionless phonons within zero ($m^0$) and first ($m^\text{I}$) approximation for the MO. In this paper, using the analytic results presented in the previous section, we shall take into account the terms of MO in the second ($m^\text{II}$) and the third ($m^\text{III}$) approximations, which essentially affect some properties of the energy spectrum. As far as there are no imaginary terms of MOs [$m_{1}(\xi), m_{2}(\xi)$], in both Green's functions [$g_{1}(\xi), g_{2}(\xi)$] all their poles are the same and, according to~(\ref{Eq4}) and (\ref{Eq6}), define the dispersion equation
\begin{displaymath}
[\xi-m_{11}(\xi)] [\xi-\delta-m_{22}(\xi)]=m_{12}^{2}(\xi), \qquad [\delta \equiv \delta_{2} = (E_{2}-E_{1})\Omega^{-1}].
\end{displaymath}
Its solutions determine all renormalized energy levels both for the main and satellite states of the system.

\begin{figure}[!t]
	\vspace{-4mm}
	\centerline{\includegraphics[width=1\textwidth]{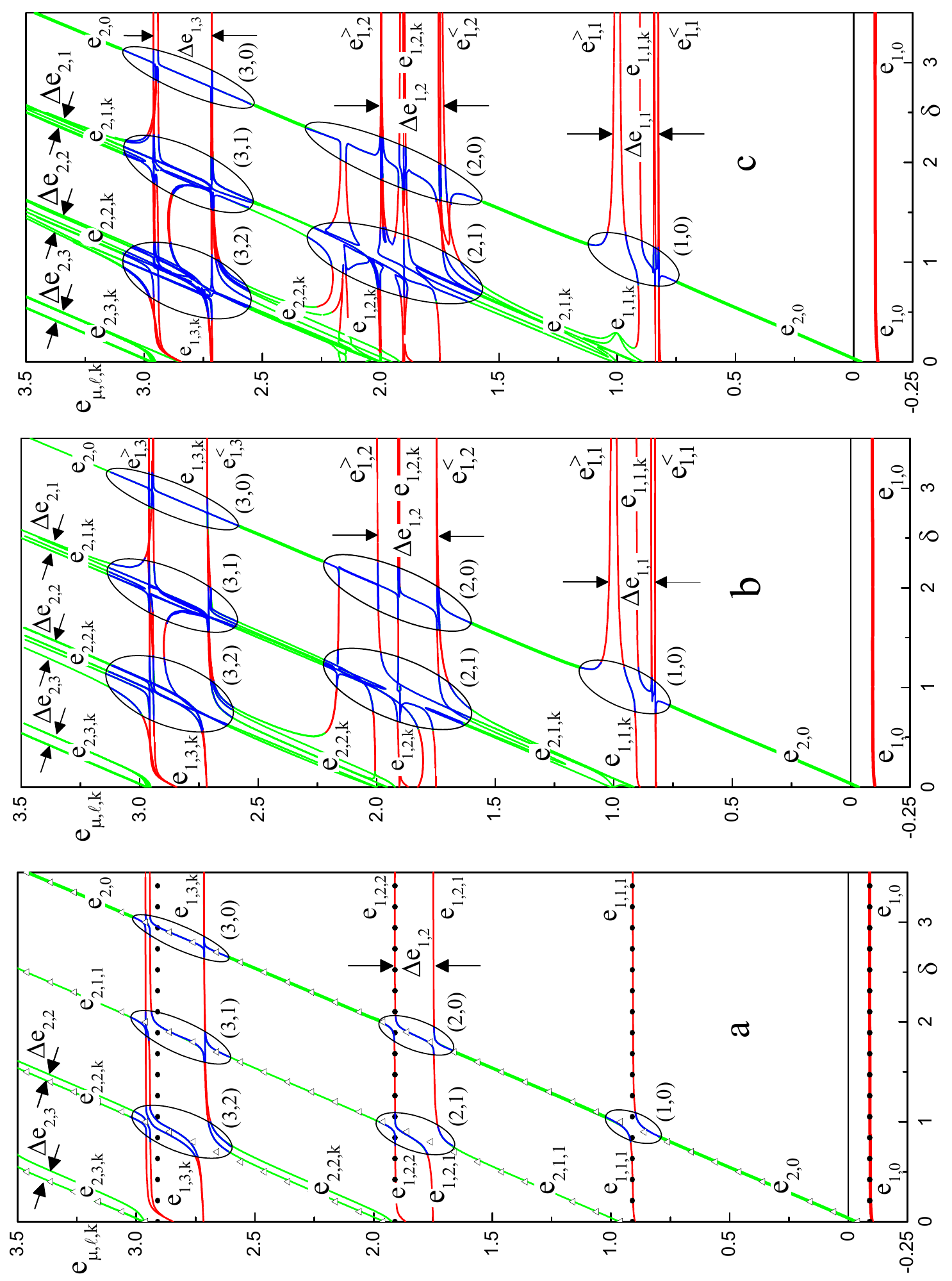}}
	\caption{(Colour online) Renormalized spectrum  $e_{\mu, l, k}$ as function of $\delta$ at $\alpha_{11}=0.3$, $\alpha_{22}=0.2$, $\alpha_{12}=0.05$.} \label{fig1}
\end{figure}

The spectrum is numerically calculated and analyzed for the system which is characterized by rather small parameters of intra-level ($\alpha_{11}=0.3, \alpha_{22}=0.2$) and inter-level ($\alpha_{12}=0.05$) coupling constants of quasiparticle and phonons. Figure~\ref{fig1} shows the spectrum obtained in four successive approximations for MO, as a function of energy distance ($\delta$) between both levels of uncoupled quasiparticle. We use convenient three-index symbols for the energy levels ($e_{\mu, l, k}$) and two-index symbols for the widths of satellite groups ($\Delta e_{\mu, l}$), which correspond to their physical origin. Thus, in $e_{\mu, l, k}$: $\mu=1, 2$ -- numerates the state of uncoupled quasiparticle, $l$ is the number of a satellite group of the levels which fixes the threshold energy of radiation accompanied by the creation of $l$ phonons, $k$  is the number of the level in the $l$-th group. The width of  $\mu l$-th satellite group ($\Delta e_{\mu, l}=e_{\mu, l}^{>}-e_{\mu, l}^{<}$) is fixed by the difference between the upper $e_{\mu, l}^{>}=e_{\mu, l, \textrm{max} k}$ and lower $e_{\mu, l}^{<}=e_{\mu, l, \textrm{min} k}$ levels of this group. Moreover, we should note that the renormalized energies of the first and second main levels are written as $e_{1,0}$ and $e_{2,0}$, while the complicated groups of levels in resonant regions are allocated by energy lines marked by two digital indices ($l_{1},l_{2}$), where $l_{1}=1,2,3 \ldots$; $l_{2}=0,1,2 \ldots$ numerate the satellite groups of the first and second main levels, respectively.

The energy levels $e_{\mu, l, k}$ calculated in different approximations for MO are presented in figure~\ref{fig1}. Herein, in the panels, the following curves correspond to: $e_{1, l, k}$ --- dots (a) $m^0$; solid red (a) $m^\text I$, (b) $m^\text{II}$, (c)~$m^\text{III}$; $e_{2, l, k}$ --- triangles (a) $m^0$; solid green (a) $m^\text I$, (b) $m^\text{II}$, (c)~$m^\text{III}$. Solid blue curves in resonant regions present the energy levels of satellite complexes which are produced by the superposition of multi-phonon states bound to the both states of uncoupled quasiparticle. Now, let us analyse how the main properties of  renormalized spectrum, calculated in different successive approximations for the MO, change.

In figure~\ref{fig1}~(a), one can see the same dependence of the spectrum on $\delta$ as obtained in approximations $m^0$ and $m^\text I$  and analyzed  in detail in our paper \cite{Tka18}. The results in these two approximations are quite different. In zero approximation ($m^0$), neglecting  the inter-level interaction ($\alpha_{12}=0$), the spectrum contains two infinite series of equidistant levels (the distance is of one-phonon energy). If $\delta$ increases, these levels intersect (except the ground one $e_{10}$) in all resonant regions of energies ($l_{1}, l_{2}$). On the contrary, in the first approximation ($m^\text I$), the anti-crossing of levels is observed in the first resonant region (1,0) while in the other resonant regions ($l_{1}=2,3,\ldots; l_{2}=0,1,\ldots$) there are multi-anti-crossings. It is due to this that in all non-resonant regions one can see two first satellite levels ($e_{1,1,1}$ and $e_{2,1,1}$) and, besides, the bands of satellite levels with the widths $\Delta e_{\mu, l=2,3,\ldots}$.

\begin{figure}[!t]
	\centerline{\includegraphics[width=0.96\textwidth]{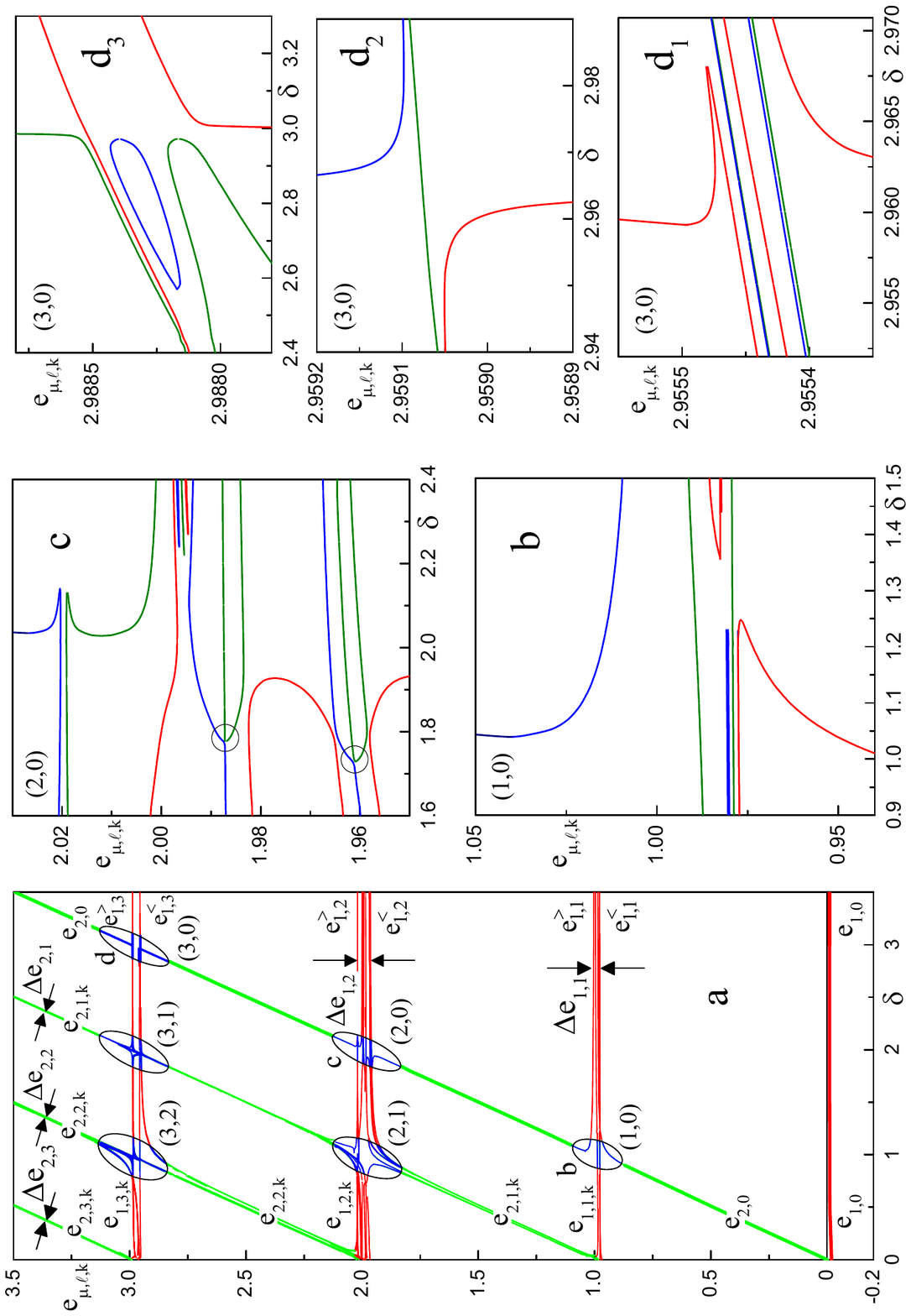}}
	\caption{(Colour online) Renormalized spectrum  $e_{\mu, l, k}$  as function of  $\delta$ at $\alpha_{11}=0.1$, $\alpha_{22}=0.08$, $\alpha_{12}=0.05$ calculated in the third approximation for the mass operator $m^\text{III}(\xi)$ (a) and bigger scales of three resonant regions (1,0) -- b, (2,0) -- c, (3,0) -- $d_{1}, d_{2}, d_{3}$.} \label{fig2}
\end{figure}

Comparing the panels (a) and (b) in figure~\ref{fig1}, one can see two qualitatively new features of the spectra calculated in the first ($m^\text I$) and second  ($m^\text{II}$) approximations for the MO. In the second approximation: i) in all resonant regions ($l_{1}, l_{2}$) there are complexes of satellite levels which do not intersect at increasing $\delta$, but some neighbouring pairs degenerate and disappear or appear. Due to the scale of figure~\ref{fig1}, this phenomenon is weakly visible, therefore, in figure~\ref{fig2} it will be shown more in detail. ii) in non-resonant regions, the number of levels in all satellite groups increases and their widths become a little bigger ($\Delta e_{\mu, l=2,3,\ldots}$). Two first satellite groups with the respective widths ($\Delta e_{1, 1}$, $\Delta e_{2, 1}$) are created instead of two first satellite levels ($e_{1, 1, 1}$ and $e_{2, 1, 1}$).

From the panels (b) and (c) in figure~\ref{fig1} it is clear that comparing the second and third approximations for the MO, the properties of the spectra are qualitatively the same. Though the number of levels in satellite complexes and groups becomes bigger, their widths increase weakly.

In figure~\ref{fig2}~(a), the spectrum calculated in the third approximation for the MO is shown for the system with smaller constants of intra-level coupling ($\alpha_{11}=0.1$, $\alpha_{22}=0.008$) but at the same magnitude of the inter-level coupling constant ($\alpha_{12}=0.005$) as for the system described in figure~\ref{fig1}. Figure~\ref{fig2}~(a) shows the same behaviour of the spectrum as figure~\ref{fig1}~(c) but here, due to the smaller $\alpha_{11}$ and $\alpha_{22}$, the sizes of the respective satellite complexes in resonant regions are smaller the same as widths of satellite bands in non-resonance regions.

For example, in the panels (b), (c), (d), figure~\ref{fig2} (with a bigger scale), the satellite levels in complexes (1,0), (2,0), (3,0), produced due to the superposition of the second main state with phonon satellites of the first main state are shown. In these panels, instead of the corresponding blue lines in figure~\ref{fig2}~(a), the lines are painted in three colors in the bottom-up sequence: red, green, blue and so on, for better visualization. In the panel (c), figure~\ref{fig2} in two circles, one can see green and blue lines which  visually seem touching each other, although in reality they are just  very close. The rest properties of the energy levels in satellite complexes are typical and well seen in the panels (a), (b), (c), (d).

\begin{figure}[!t]
\centerline{\includegraphics[width=1\textwidth]{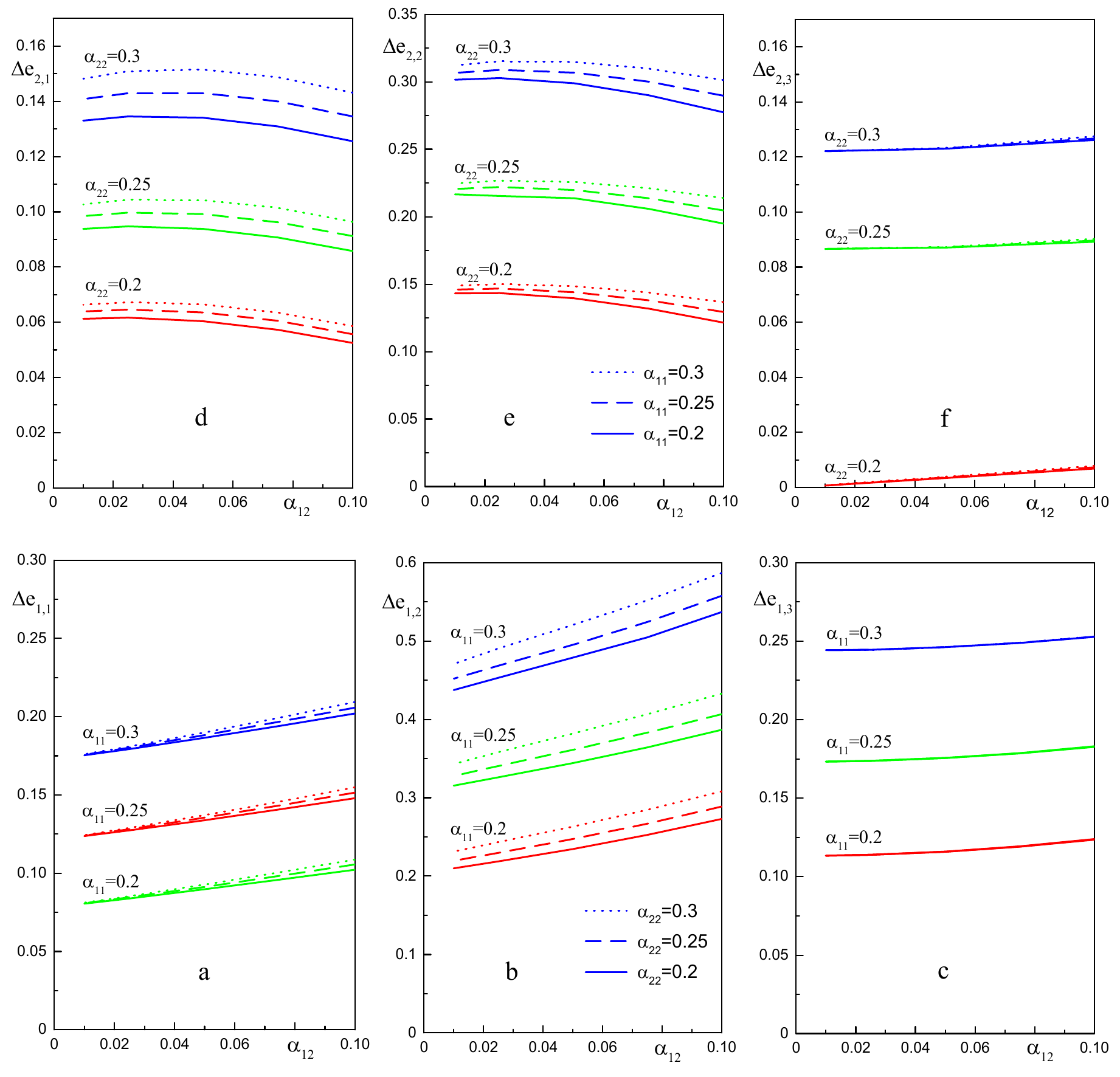}}
\caption{(Colour online) Widths of satellite bands ($\Delta e_{\mu, l}$) as functions of the inter-level coupling constant ($\alpha_{12}$) at different intra-level coupling constants ($\alpha_{11}, \alpha_{22}$), presented in the panels.} \label{fig3}
\end{figure}

The widths of satellite bands are important parameters of the spectrum in the non-resonant energy region, depending on intra- and inter-level coupling constants. The results of their calculations in the third approximation are presented in figure~\ref{fig3}, with $\alpha_{11}$, $\alpha_{22}$, $\alpha_{12}$  values at the respective panels. The figure proves that the widths of all satellite bands quasi-linearly depend on the magnitude of inter-level coupling in the range $0 < \alpha_{12} \leqslant 0.1$. The panels (a), (b), (c) show that the widths of three satellite bands of the first level ($\Delta e_{1, l}$) are mainly determined by the magnitude of intra-level coupling ($\alpha_{11}$) of quasiparticle in the same state with phonons, while increasing $\alpha_{22}$ or $\alpha_{12}$  weakly increases the widths  $\Delta e_{1, 1}$  and $\Delta e_{1, 2}$. The width $\Delta e_{1, 3}$ almost does not depend on $\alpha_{22}$ and $\alpha_{12}$. The widths of three satellite bands of the second level ($\Delta e_{2, l}$), panels (d), (e), (f), are mainly determined by $\alpha_{22}$ magnitude while $\alpha_{11}$ and $\alpha_{12}$ affect weakly.

\section{Main results and conclusions}
To calculate the energy spectrum of multi-level quasiparticle interacting with polarization phonons described by Fr\"ohlich-type Hamiltonian at $T=0$~K, the Feynman-Pines diagram technique is generalized for an effective renormalization of mass operator due to multi-phonon processes.

Within the approach of successive separation of infinite classes of multiplicative diagrams from the blocks of non-multiplicative diagrams of MO and their partial summing, the MO is finally expressed as the sum of continuous branch fractions with typical links. Such a representation effectively takes into account multi-phonon processes and avoids the so-called ``problem of sign''.

The renormalized energy spectrum is calculated using  four successive approximations for the MO within the example of a two-level quasiparticle interacting with phonons. The analysis of this spectrum in different approximations shows that its properties essentially depend on the inter-level interaction.

In zero approximation for the MO (neglecting  inter-level interaction), the spectrum contains two infinite series of equidistant satellite levels, which intersect in all resonant regions  when the energy distance between two levels of uncoupled quasiparticle varies. In the first and further approximations for MO, instead of two series of single satellite levels, the complexes of satellite levels are observed in resonant regions while the bands of satellite levels are observed in non-resonant regions. In higher approximations, the number of levels in satellite complexes and bands becomes bigger though their widths vary weakly. At increasing $\delta$, the levels do not intersect but some neighbouring levels degenerate (disappear or appear), which causes an unstable spectrum. It is shown that the widths of satellite bands are mainly proportional to the magnitudes of the intra-coupling constants ($\alpha_{\mu\mu}$) of those main levels~($\mu$) of an uncoupled quasiparticle, by which they are produced due to the interaction with phonons. The sizes of satellite complexes are characterized by both widths of  those two bands that produce them by intersecting.

Observing the system under research as the simplest (rough) model of electron-phonon interaction in basic elements of modern nano-devices (QCD and QCL), one can expect that the revealed bands and complexes of satellite levels (which were not taken into account in the previous theoretical papers) would essentially affect their operation. In particular, the satellite complexes and bands of the levels can replace the missing ``steps'' of the ``torn phonon ladder'' or imperfect ``complete phonon ladder'' in extractors of QCD operating in IR-range.

Of course, for a further development of a consistent theory of electron-phonon interaction in multi-layered  nanostructures, due to the effect of spatial quantization, it is necessary to take into account both the multi-band quasiparticles and all modes of the phonon spectrum that are present in real structures. This will generally make the theory much more complicated, but will make it possible to make a research into the new interesting and important properties and physical phenomena in nanoheterostructures.

\appendix
\section{MO matrix elements $m^{g(\tau)}_{\mu
		\mu}$ at ($\tau = 2,3,4,5$)}
\label{appendix_A}

\begin{equation}
m_{11}^{g (2)}=m_{12} m_{21}  \bar{\xi}_2^{\ -1},
\end{equation}

\begin{equation}
m_{11}^{g (3)}=\big(m_{12}^2 \bar{\xi}_3 + m_{13}^2 \bar{\xi}_2 + 2 m_{12} m_{23} m_{31} \big) \left(\bar{\xi}_2 \bar{\xi}_3 - m_{23} m_{32}\right)^{-1},
\end{equation}

\begin{equation}
\begin{array}{c}
m_{11}^{g (4)}=\Big\{ m_{12}^2 \left(\bar{\xi}_3 \bar{\xi}_4 - m_{34}^2 \right) + m_{13}^2 \left(\bar{\xi}_2 \bar{\xi}_4 - m_{24}^2 \right)+ m_{14}^2 \left(\bar{\xi}_2 \bar{\xi}_3 - m_{23}^2 \right) \\
\\
+ 2 \left[m_{12} m_{13} \left(m_{23}\bar{\xi}_4 + m_{24} m_{34} \right) + m_{12} m_{14} \left(m_{24} \bar{\xi}_3 + m_{23} m_{43} \right) + m_{13} m_{14} \left(m_{34} \bar{\xi}_2 + m_{32} m_{42} \right) \right] \Big\} \\
\\
\times\left(\bar{\xi}_2 \bar{\xi}_3 \bar{\xi}_4 - m_{23}^2 \bar{\xi}_4 - m_{34}^2 \bar{\xi}_2 - m_{42}^2 \bar{\xi}_3 - 2 m_{23} m_{34} m_{42}\right)^{-1},
\end{array}
\end{equation}

\begin{equation*}
\begin{array}{c}
m_{11}^{g (5)}=\Big\{m_{12}^2\big(\bar{\xi}_3\bar{\xi}_4\bar{\xi}_5-m_{34}^2\bar{\xi}_5-m_{45}^2\bar{\xi}_3-m_{53}^2\bar{\xi}_4-2 m_{34} m_{45} m_{53}\big)\\
\\
+m_{13}^2\big(\bar{\xi}_2\bar{\xi}_4\bar{\xi}_5-m_{24}^2\bar{\xi}_5-m_{45}^2\bar{\xi}_2-m_{52}^2\bar{\xi}_4-2 m_{24} m_{45} m_{52}\big)\\
\\
+m_{14}^2\big(\bar{\xi}_2\bar{\xi}_3\bar{\xi}_5-m_{23}^2\bar{\xi}_5-m_{35}^2\bar{\xi}_2-m_{52}^2\bar{\xi}_3-2 m_{23} m_{35} m_{52}\big)\\
\\
+m_{15}^2\left(\bar{\xi}_2\bar{\xi}_3\bar{\xi}_4-m_{23}^2\bar{\xi}_4-m_{34}^2\bar{\xi}_2-m_{42}^2\bar{\xi}_3-2 m_{23} m_{34} m_{42}\right)\\
\\
+2m_{12}m_{13}\left[m_{23}\big(\bar{\xi}_4\bar{\xi}_5-m_{45}^2\big)+m_{24}m_{34}\bar{\xi}_5+m_{25}m_{35}\bar{\xi}_4+m_{45} \left(m_{25} m_{34}+m_{24} m_{35}\right)\right]\\
\\
+2m_{12}m_{14}\left[m_{24}\big(\bar{\xi}_3\bar{\xi}_5-m_{35}^2\big)+m_{23}m_{43}\bar{\xi}_5+m_{25}m_{45}\bar{\xi}_3+m_{35} \left(m_{25} m_{43}+m_{23} m_{45}\right)\right]\\
\\
+2m_{12}m_{15}\left[m_{25}\big(\bar{\xi}_3\bar{\xi}_4-m_{34}^2\big)+m_{23}m_{53}\bar{\xi}_4+m_{24}m_{54}\bar{\xi}_3+m_{34} \left(m_{24} m_{53}+m_{23} m_{54}\right)\right]\\
\\
+2m_{13}m_{14}\left[m_{34}\big(\bar{\xi}_2\bar{\xi}_5-m_{25}^2\big)+m_{32}m_{42}\bar{\xi}_5+m_{35}m_{45}\bar{\xi}_2+m_{25} \left(m_{35} m_{42}+m_{32} m_{45}\right)\right]\\
\\
+2m_{13}m_{15}\left[m_{35}\big(\bar{\xi}_2\bar{\xi}_4-m_{24}^2\big)+m_{32}m_{52}\bar{\xi}_4+m_{34}m_{54}\bar{\xi}_2+m_{24} \left(m_{34} m_{52}+m_{32} m_{54}\right)\right]\\
\\
+2m_{14}m_{15}\left[m_{45}\big(\bar{\xi}_2\bar{\xi}_3-m_{23}^2\big)+m_{42}m_{52}\bar{\xi}_3+m_{43}m_{53}\bar{\xi}_2+m_{23} \left(m_{43} m_{52}+m_{42} m_{53}\right)\right]\Big\} \\
\end{array}
\end{equation*}
\begin{equation}
\begin{array}{c}
\times\Big\{\bar{\xi}_2\bar{\xi}_3\bar{\xi}_4\bar{\xi}_5-m_{23}^2\bar{\xi}_4\bar{\xi}_5-m_{24}^2\bar{\xi}_3\bar{\xi}_5-m_{25}^2\bar{\xi}_3\bar{\xi}_4-
m_{34}^2\bar{\xi}_2\bar{\xi}_5-m_{35}^2\bar{\xi}_2\bar{\xi}_4-m_{45}^2\bar{\xi}_2\bar{\xi}_3\\
\\
-2\left[m_{23}m_{34}\left(m_{24}\bar{\xi}_5+m_{25} m_{45}\right)+m_{24}m_{45}\left(m_{25}\bar{\xi}_3+m_{23} m_{53}\right)+m_{25}m_{53}\left(m_{32}\bar{\xi}_4+m_{24} m_{34}\right)+m_{34}m_{45}m_{53}\bar{\xi}_2\right]\\
\\
+m_{23}^2 m_{45}^2+m_{24}^2 m_{53}^2+m_{25}^2 m_{34}^2\Big\}^{-1},
\end{array}
\end{equation}
where $\bar{\xi}_ {\mu}=\xi_{\mu} - m_{\mu \mu}(\xi)$.

\ukrainianpart

\title{Метод послідовного виділення та підсумовування мультиплікативних діаграм масового оператора для систем багаторівневих квазічастинок взаємодіючих з поляризаційними фононами}
\author{М.В. Ткач, Ю.О. Сеті, О.М. Войцехівська, В.В. Гутів}
\address{Чернівецький національний університет ім. Ю. Федьковича, \\ вул. Коцюбинського, 2,  58012 Чернівці, Україна}

\makeukrtitle

\begin{abstract}
\tolerance=3000%
На основі діаграмної техніки Фейнмана-Пайнса у новому підході розвинена теорія перенормованого спектра систем багаторівневих квазічастинок взаємодіючих з поляризаційними фононами при $T=0$~K, який дозволяє послідовно виділяти мультиплікативні з немультиплікативних діаграм усіх порядків масового оператора і здійснювати їх парціальне підсумовування. Отриманий у цьому підході масовий оператор є сумою розгалужених ланцюгових дробів, які ефективно враховують багатофононні процеси.
На прикладі системи дворівневих квазічастинок показано, що саме міжрівнева (недіагональна) взаємодія з фононами кардинально змінює властивості перенормованого спектра. У околах усіх порогових енергій утворюються квазіеквідистантні фононні сателітні зони --- групи енергетичних рівнів, які відповідають комплексам зв’язаних станів квазічастинки з багатьма фононами.

\keywords діаграмна техніка, квазічастинка, масовий оператор, фонон, спектр

\end{abstract}

\end{document}